\documentclass[sigconf]{acmart}
\acmConference[FORGE 2026]{3rd ACM International Conference on AI Foundation Models and Software Engineering}{April 2026}{Rio de Janeiro, Brazil}

\usepackage{graphicx} 
\usepackage{url} 
\usepackage{xcolor} 
\usepackage{hyperref}
\usepackage{placeins}
\usepackage{float}
\usepackage{stfloats}

\title{SEMODS: A Validated Dataset of Open-Source \underline{S}oftware \underline{E}ngineering \underline{Mod}el\underline{s}}

\author{Alexandra González, Xavier Franch, Silverio Martínez-Fernández}
\affiliation{
  \institution{Universitat Politècnica de Catalunya}
  \city{Barcelona}
  \country{Spain}}
\email{{alexandra.gonzalez.alvarez, xavier.franch, silverio.martinez}@upc.edu}

\newcommand{\ourDS}{SEMODS}

\begin{abstract} 
Integrating Artificial Intelligence into Software Engineering (SE) requires having a curated collection of models suited to SE tasks. With millions of models hosted on Hugging Face (HF) and new ones continuously being created, it is infeasible to identify SE models without a dedicated catalogue. To address this gap, we present \ourDS: an SE-focused dataset of 3,427 models extracted from HF, combining automated collection with rigorous validation through manual annotation and large language model assistance. Our dataset links models to SE tasks and activities from the software development lifecycle, offering a standardized representation of their evaluation results, and supporting multiple applications such as data analysis, model discovery, benchmarking, and model adaptation. 
\end{abstract} 


\begin{document}

\maketitle

\section{Introduction}
Pre-Trained Models (PTMs) represent deep neural network architectures that have been trained on a specific dataset using well-defined data processing and training pipelines, resulting in learned model parameters (weights) \cite{davis2023reusing}. Within software projects, PTMs may serve as core components or be used for experimentation \cite{koohjani2025exploring}. 

PTMs are shared through Machine Learning (ML) registries, also known as model hubs or model zoos, where teams collaborate and share ML assets
\cite{jiang2023empirical, jiang2024peatmoss}. 
These registries play a key role in fostering reuse, as they reduce the cost and effort associated with training models from scratch while promoting reproducibility \cite{davis2023reusing}. 

Hugging Face (HF) \cite{huggingface} is a popular open-source registry designed for sharing and developing models, datasets, and applications built with them (known as spaces). Each model repository in HF stores a rich set of attributes, ranging from popularity metrics (e.g., likes, downloads) to metadata such as licenses, libraries, training datasets, and inference providers. Owing to its openness and active community, the platform has experienced remarkable growth. During 2024 alone, HF recorded an average of 2,199 new models created per day and over six million daily downloads, according to our own calculations. As of November 2025, the platform hosts over two million models \cite{laufer2025anatomy} and continues to grow rapidly, with a new repository created approximately every fifteen seconds \cite{linkedinJustCrossed}. 

Despite this wealth of resources, SE researchers \cite{giray2021software} and practitioners \cite{tan2024challenges,zhao2024empirical, ajibode2025towards} still struggle to identify models that are directly relevant to their tasks, while satisfying project constraints such as licensing or performance. Common issues include missing attributes, discrepancies between reported and actual performance, and risks related to privacy or unethical model behaviour \cite{jiang2023empirical}. The lack of an SE-focused catalogue in ML registries \cite{10.1145/3661167.3661215} limits the integration of ML into the Software Development Life Cycle (SDLC) \cite{sommerville2015}, forcing users to manually navigate vast collections of models to locate suitable candidates. This process is time-consuming and error-prone, ultimately slowing the adoption of ML in SE workflows \cite{8804457}. 

To address this gap, we present \underline{S}oftware \underline{E}ngineering \underline{Mod}el\underline{s} (\ourDS): a dataset of SE models, systematically collected, processed, and validated to support their in-depth analysis, efficient discovery, and practical use within SE contexts. The dataset enables researchers and practitioners to explore models tailored to specific SE tasks and interact with their associated metadata through SQL queries, while also supporting benchmarking, and facilitating the identification of models for model adaptation.

The contributions of this work are as follows. 
\begin{enumerate}
    \item A validated dataset of 3,427 HF SE models, catalogued according to SE activities and tasks across the SDLC.
    \item A standardized representation of benchmarks and metrics.
    \item Automate processes that keep the dataset up to date with new HF repositories and refresh dynamic attributes.
\end{enumerate}

\textbf{Data availability}: A snapshot of the dataset (November 2025 release), containing 3,427 SE models and their associated attributes, is publicly available in Zenodo \cite{zenodoSEMODSValidated}. To foster reproducibility, the full cataloguing pipeline is also accessible in Zenodo \cite{anonymous_author_s_2025_17674909}.

\section{Related Work}
Several efforts have collected and organized data from model registries to enable the reuse and large-scale analysis of models, including those relevant to SE. 

Ait et al. \cite{10123660} introduced HFCommunity, a relational database that aggregates data from the HF Hub. This database was developed to address the absence of tools for collecting and exploring HF data beyond the platform's API. 

Similarly, Jiang et al. \cite{jiang2023ptmtorrent} proposed PTMTorrent, a dataset designed to facilitate the evaluation and understanding of PTM packages, including pre-trained weights, documentation, model architectures, datasets, and metadata. PTMTorrent consolidates information from five model hubs (HF, Model Zoo, PyTorch Hub, ONNX Model Zoo), covering 15,913 packages. Due to space constraints, the HF subset only comprises the 10\% most-downloaded models. 

In a subsequent effort, Jiang et al. \cite{jiang2024peatmoss} released the PeaTMOSS dataset, which comprises metadata for 281,638 PTMs and detailed snapshots for those with over 50 monthly downloads (14,296 PTMs). PeaTMOSS also connects PTMs with 28,575 open-source software repositories from GitHub that use them, establishing 44,337 mappings between 15,129 downstream GitHub repositories and the corresponding 2,530 PTMs. This integration of model and repository data enables opportunities for mining PTMs and investigating the PTM supply chain. 

Compared to prior work, we have specifically curated and validated models with explicit relevance to SE, introducing novel mappings to SE tasks and standardized benchmarking data. Our dataset provides a way for users to efficiently discover and access SE models relevant to their tasks. This work extends existing datasets, such as HFCommunity \cite{10123660} and PeaTMOSS \cite{jiang2024peatmoss}, by incorporating the SE-focus catalogue and a harmonization of the evaluation information, enabling targeted exploration and reuse of PTMs in the SDLC.

\section{Dataset Construction and Applications}
\begin{figure}[!h]
    \centering
    \includegraphics[width=1\linewidth]{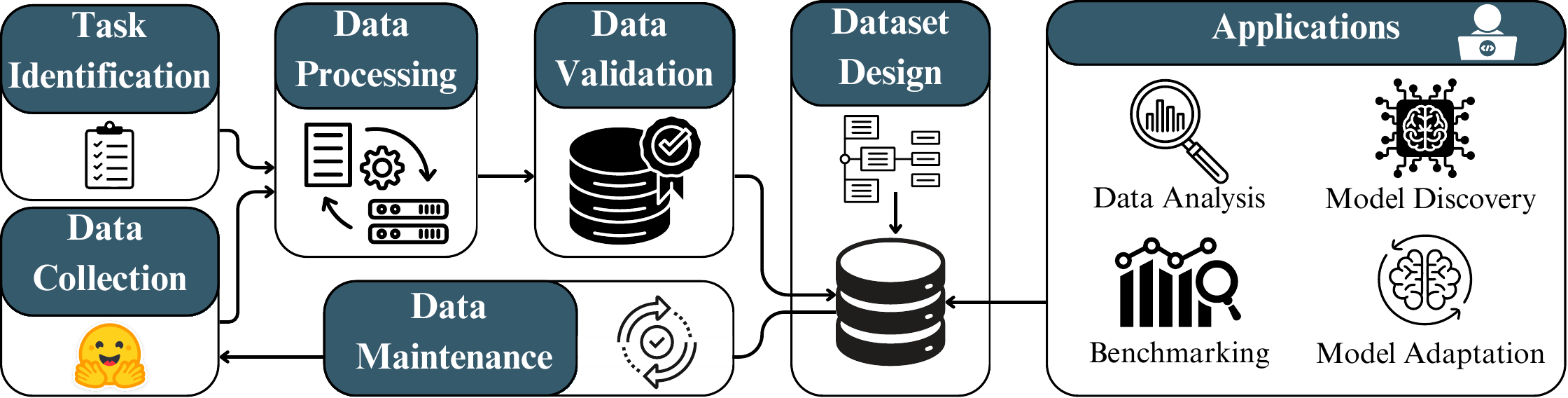}
    \caption{Dataset construction process and applications.}
    \label{fig:schema}
\end{figure}

Figure~\ref{fig:schema} illustrates the process used to build the dataset and indicates its applications. Below, we detail the cataloguing process, consisting of task identification, data collection, processing, and validation, followed by a description of the dataset contents through its conceptual schema. Next, we describe the maintenance practices that keep the dataset up to date by monitoring new models and refreshing dynamic attributes, and we discuss its main applications. The cataloguing pipeline was first executed on a complete snapshot of HF models as of March 2025, encompassing 1.5 million models, and once validated, it became an automated workflow that applies the same steps to newly released assets on a daily basis. 

\subsection{Task Identification}\label{subsec:task_identification}
We derived a taxonomy of SE tasks across the five SDLC stages by building on prior work. Starting from the 88 tasks identified by Hou et al. \cite{10.1145/3695988}, we aligned them with established sources such as Sommerville \cite{sommerville2015}, including renaming software development activity to software implementation. We further refined the taxonomy by treating software architecture as part of software design, and by adding tasks guided by the chapters of Software Engineering Body of Knowledge \cite{washizaki2024swebok}. The resulting taxonomy, refined through discussion among the authors to ensure coverage and avoid overlap, comprises 147 tasks. We note that the current set of HF models covers 100 SE tasks, but as the cataloguing pipeline runs daily, coverage may increase over time as new models are published.

\subsection{Data Collection}\label{subsec:data_coll}
We retrieved all models hosted on HF via its API \cite{huggingfaceHuggingFace}. For each asset, we considered all available documentation associated with it. Specifically, we collected, whenever it existed: (i) its \textit{model card description}, a markdown file documenting the model's characteristics, intended uses, and evaluation \cite{mitchell2019model}; (ii) the associated \textit{card metadata}, specified in a YAML block (e.g., license, tags, language) \cite{huggingfaceModelCardsMetadata}; and (iii) the \textit{abstract} of a linked arXiv paper \cite{huggingfaceModelCardsPaper}, taking advantage of the cross-platform linking between HF and arXiv \cite{suryani2024exploration}.

\subsection{Data Processing}\label{subsec:data_proc}
After collecting the data, we processed it to facilitate automatic cataloguing. We began with text normalization, including tokenization, lowercasing, and lemmatization, to enable accurate detection of SE tasks within the model documentation. As we focused on SE-relevant resources, we searched for SE tasks in the processed text, requiring multi-word tasks (e.g., ``code generation") to appear with all words together and in the correct order, and rejecting partial matches that only appeared inside longer tokens.
To ensure the rigour of our matches, we identified outliers (i.e., SE tasks with abnormally high frequencies) and unique instances arising from high textual similarity between model documentation entries, thereby preventing duplicates or multiple counts of the same PTM.

\subsection{Data Validation}\label{subsec:data_val}
Lastly, we conducted a rigorous validation to ensure that only SE-relevant resources were retained, combining manual annotation with Large Language Model (LLM) assistance in a two-phase validation process. In the first phase, we randomly selected a subset of models for each SE activity, ensuring coverage across all associated tasks. Sample sizes were calculated using a 95\% confidence level and a 5\% margin of error to ensure statistical validity \cite{qualtricsSampleSize}. The first author manually annotated all these subsets, totalling 1,346 models. Following established interceder reliability practices in qualitative research \cite{o2020intercoder}, two other authors with experience in the domain annotated 10\% of the samples for each SE activity. The resulting annotated data served as ground truth for the LLM (Gemini 2.0 Flash), which was prompted in a zero-shot setting to provide both a binary relevance judgment and a rationale for each PTM in the pilot set. After assessing the model’s performance using Cohen’s kappa \cite{sim2005kappa} and refining all five prompts until we obtained an almost perfect level of agreement ($k>0.8$), a second validation phase tested the LLM’s generalization on previously unseen data. Once these tests confirmed the model’s reliability, we used the LLM to determine whether each model in the full set addressed an SE activity.

\subsection{Dataset Design} 

\begin{figure*}[!ht] 
    \centering
    \includegraphics[width=0.9\linewidth]{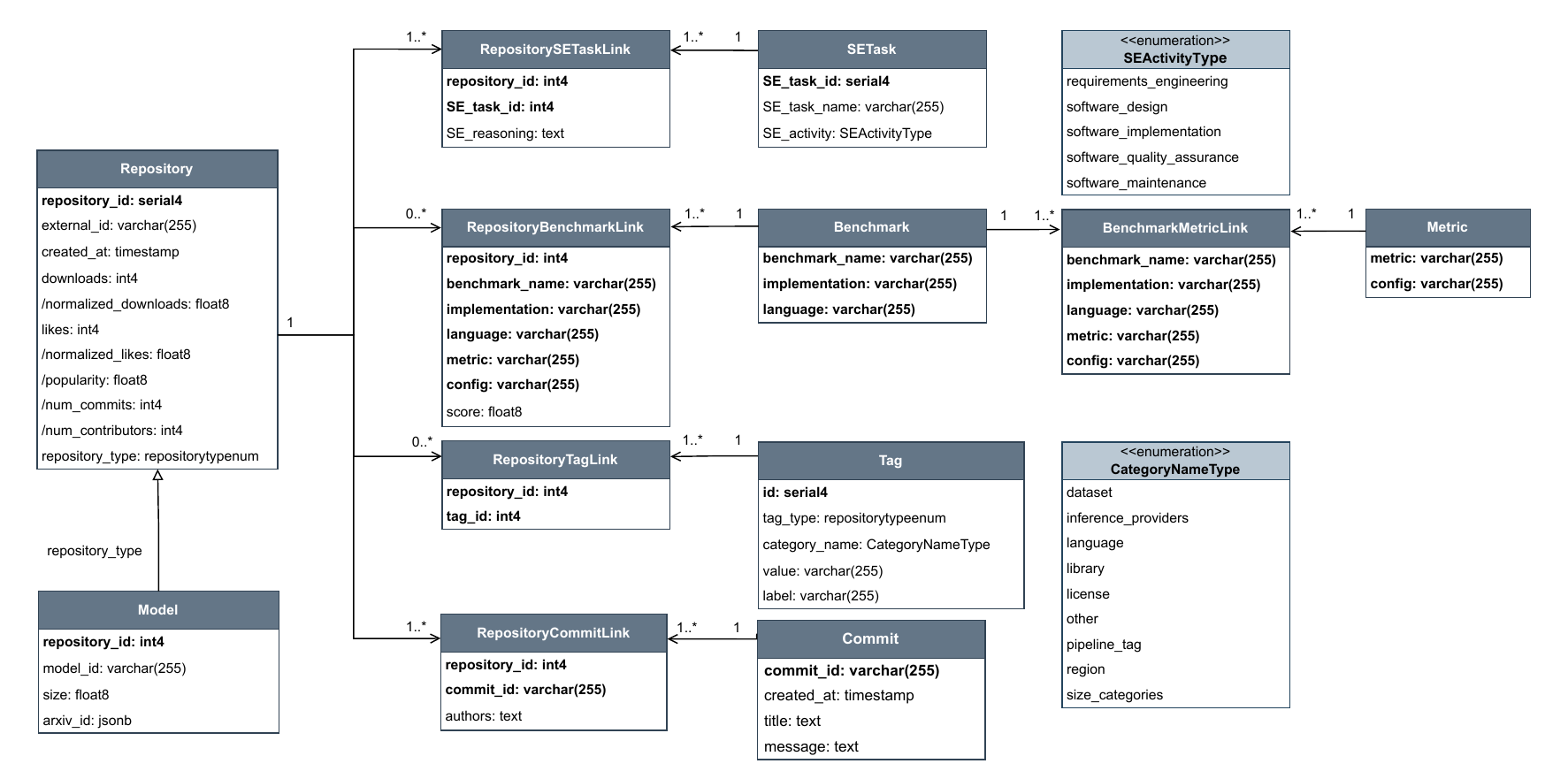}
    \caption{Conceptual schema of the dataset. This design introduces novel entities specific to SE models (e.g., SETask) and their standardized evaluation tables (e.g., Benchmark, Metric), enabling the development and assessment of SE models.}
    \label{fig:uml}
\end{figure*}

The dataset schema is defined as a UML class diagram, shown in Figure~\ref{fig:uml}. This representation structures HF repositories from an SE perspective, enabling querying and facilitating the analysis and effective use of the ML ecosystem in SE.

A hierarchy represents repository types through the \texttt{Repository} class and its subclass \texttt{Model}, which specifically represents models. To allow for future extensions of the dataset to include other HF categories besides models (such as datasets or spaces), the \texttt{Repository} class includes the attribute \texttt{repository\_type}, which specifies the repository category in HF. Each \texttt{repository} instance stores identifiable metadata (our internal \texttt{repository\_id} and the \texttt{external\_id} defined by HF), and descriptive attributes characterizing HF repositories. These include temporal information (\texttt{created\_at}), popularity indicators (original and normalized values for \texttt{downloads} and \texttt{likes}, along with the derived \texttt{popularity}), and activity metrics (\texttt{num\_commits} and \texttt{num\_contributors}). The \texttt{Model} subclass extends this information with attributes such as \texttt{arxiv} for paper identifiers, and \texttt{size}. Since the platform does not directly provide model size, we compute it in bytes based on the total size of the files associated with each model.

Repositories are connected to four entities that describe their characteristics. The first is the \texttt{SETask} class, which specifies the SE activity (\texttt{activity}) and task (\texttt{name}) that the PTM supports. The possible values for the \texttt{activity} attribute are closed and correspond directly to the five stages of the SDLC (\texttt{SEActivityType}). Additionally, a single model is not necessarily constrained to one specific SE task or activity, as many models have a broad utility across the lifecycle. This mapping between HF repositories and specific SE activities and tasks, absent in the current HF metadata, is derived from the cataloguing process detailed in the dataset construction's earlier steps. This information is particularly useful for integrating ML assets into SE, as it directly maps the model’s utility to specific parts of the SDLC. Finally, the \texttt{RepositorySETaskLink} class formally links repositories with their detected SE tasks and includes a \texttt{reasoning} attribute describing the evidence extracted from the model documentation. 

The second descriptive entity is the \texttt{Benchmark} class, which contains structured evaluation information for each repository through the \texttt{RepositoryBenchmarkLink} class. Although model cards provide results in the \texttt{model-index} section, the information is heterogeneous and lacks standard reporting. To address this, we applied a rule-based normalization to standardize the inconsistent naming in the HF documentation (e.g., unifying benchmark variants such as \textit{MMLU} and \textit{Measuring Massive Multitask Language Understanding}, or metrics such as \textit{accuracy\_norm} and \textit{normalized\_accuracy}). For each repository with available evaluation data, we obtained the \texttt{benchmark\_name}, the \texttt{implementation} framework, the programming \texttt{language}, the performance \texttt{metric}, its configuration (\texttt{config}), and the corresponding numeric \texttt{score}. We obtained a set of 206 distinct benchmarks (e.g., HumanEval \cite{chen2021evaluating}, MBPP \cite{austin2021program}) and 43 different metrics (e.g., normalized accuracy, cosine accuracy). 

The third entity is the \texttt{Tag} class, a key feature in HF that provides additional information associated with the PTM \cite{suryani2025model}. Each tag instance specifies its category (\texttt{category\_name}), a descriptive \texttt{label}, its type (\texttt{tag\_type}), and a corresponding \texttt{value}. Tag categories encompass a wide range of repository attributes, including the training \texttt{dataset}, the available \texttt{inference\_providers}, supported \texttt{language}, underlying \texttt{library}, \texttt{license}, associated \texttt{pipeline\_tag}, deployment \texttt{region}, and \texttt{size\_categories}, among others. Tags are linked to repositories through the \texttt{RepositoryTagLink} class, allowing each repository to be associated with multiple tags.

Finally, we have the \texttt{Commit} class. As the fourth descriptive entity, it captures information about the changes made to the repository. The \texttt{Commit} class includes a unique identifier (commit\_id), temporal information (\texttt{created\_at}), and text fields that store the commit's details, such as the \texttt{title}, and \texttt{message}. Repositories are linked to these records through the \texttt{RepositoryCommitLink} class, which also includes an author attribute to document who made the change. Given that models are often reused or forked, a single \texttt{Commit} can be associated with multiple repositories via this link table.

Using the collected, processed, and validated data along with the conceptual schema, we generated \ourDS{} tables. Two auxiliary tables support this process: one stores the raw repositories retrieved from the HF API, and the other contains only SE repositories. 

\subsection{Data Maintenance}
Given the fast-growing pace of the HF ecosystem \cite{han2021pre,ait2024hfcommunity} and the increasing rate of new repositories and model contributions \cite{10.1145/3643991.3644898}, we implemented an automated process that checks daily for newly published HF models. Each new repository is automatically catalogued or discarded if deemed irrelevant to SE. In addition, dynamic attributes (e.g., \texttt{likes}, \texttt{downloads}, \texttt{num\_commits}, \texttt{num\_contributors}) are refreshed twice per day to maintain accurate metrics.

\subsection{Applications}
The relational structure and SE-specific focus of \ourDS{} opens multiple opportunities for researchers and practitioners working on Artificial Intelligence (AI) for SE. The data enables new empirical studies and supports practical use cases for integrating AI into the SDLC. Below, we outline its main applications along with example Research Questions (RQs) that can be addressed using the data.

\subsubsection{Data Analysis}
\ourDS{} supports quantitative and qualitative analyses of SE models in HF, enabling the characterization of the ML ecosystem from an SE perspective. Quantitative analyses can reveal trends in model creation and reuse, while qualitative exploration can examine repository documentation or benchmarking practices. The availability of structured evaluation data enables insights, such as correlations between model performance, size, and popularity. Cross-referencing \ourDS{} with generalist datasets highlights its specialized scope: only 244 (7.12\%) and 990 (28.89\%) of our SE models overlap with PeaTMOSS \cite{jiang2024peatmoss} and HFCommunity \cite{10123660}, respectively. This minimal overlap (<0.2\% of those collections) stems from \ourDS' recency (PeaTMOSS has been static since August 2023, HFCommunity since October 2024) and inclusivity (\ourDS{} retains specialized and emerging models regardless of popularity). These analyses support RQs such as: \textit{Who creates SE models in HF (e.g., newcomers or experienced developers)?} \textit{How does model maintenance health relate to sustained popularity in SE models?}  \textit{Which benchmarks are used to evaluate SE models?}

\subsubsection{Model Discovery}
Users can query \ourDS{} to find models for their integration into SE pipelines using SE-specific attributes (e.g., SE task, SE activity). For instance, users may search for models supporting ``software design'' or ``code summarization'' tasks while also meeting open policy standards or metric constraints. The inclusion of the \texttt{reasoning} attribute promotes transparency on how each PTM maps to an SE activity, supporting semantic, metric-based, and learning-based selection methods \cite{zhouunifying}, and reducing manual exploration at scale. This enables questions such as: \textit{Which SE activities are more and least supported by existing models?} \textit{How do discovery patterns vary when filtering by specific attributes?}

\subsubsection{Benchmarking}
By providing structured evaluation results across heterogeneous benchmarks and metrics, users can conduct empirical studies on PTM performance. Researchers can perform cross-benchmark comparisons to investigate performance variability, and compare results for a given benchmark and configuration to identify the best-performing models. Moreover, the data can reveal which benchmarks are commonly used for specific SE activities and tasks, highlighting underexplored areas where new benchmarks may be needed. These capabilities support RQs such as: \textit{How consistent are model results across benchmarks?} \textit{Which SE activities lack systematic evaluation resources or need new ones?}

\subsubsection{Model Adaptation}
As each PTM is associated with multiple attributes, the dataset facilitates the identification of candidate models for adaptation through approaches like fine-tuning, transfer learning \cite{zhuang2020comprehensive}, and knowledge distillation \cite{zhouunifying, davis2023reusing}. Users can identify models trained on similar datasets or SE tasks, providing a starting point for developing SE-specific models. For example, developers seeking to fine-tune a model for \textit{code editing} can explore metadata on existing models trained on \textit{code generation} to select a suitable candidate, thereby reducing training time, computational costs, and energy consumption. These attributes open the door to RQs such as: \textit{Which models are best suited for adaptation to specific SE tasks?} \textit{How does training dataset similarity influence adaptation outcomes?}

\section{Threats to Validity}
We acknowledge some threats to the validity of \ourDS. Internal validity is at risk for models with poor or missing documentation, while models with complete model cards are well supported. We mitigated this by enriching our data collection process with external evidence, such as linked arXiv abstracts. External validity is limited to the HF Hub and subject to future changes in HF’s documentation practices. The construct validity relies on our taxonomy of SE tasks, which builds upon established literature \cite{10.1145/3695988, washizaki2024swebok}. Finally, regarding conclusion validity, we addressed the risk of using an LLM to confirm SE relevance by designing a validation protocol involving three independent human annotators (see Section \ref{subsec:data_val}).

\section{Conclusions and Future Work}
We have presented \ourDS, a dataset comprising information about 3,427 SE models in HF. The data was collected using a rigorous pipeline that scans all models available in the platform and catalogues them according to the SDLC. 
Beyond the metadata already provided in HF, our dataset introduces novel mappings of models to specific SE tasks and activities, as well as a standardized representation of the reporting evaluation metrics by extracting and harmonizing benchmark and metric configurations. 

In future work, we plan to expand the dataset with additional sources of information (e.g., GitHub Models \cite{githubGitHubModels}, PyTorch Hub \cite{pytorchPyTorch}). We also intend to combine our dataset with existing ones that provide complementary attributes, and to develop a recommender system that assists users in identifying models of potential interest.

\section{Acknowledgments}
This work was supported by Grant PID2024-156019OB-I00 funded by MICIU/AEI/10.13039/501100011033 and by ERDF, EU. Alexandra González additionally thanks the FI-STEP grant 2025 STEP-00407.

\balance
\bibliographystyle{acm}
\bibliography{references}

@misc{huggingface,
	author = {{Hugging Face}},
	title = {{H}ugging {F}ace – {T}he {A}{I} community building the future. --- huggingface.co},
	howpublished = {\url{https://huggingface.co}},
	year   = {[n.d.]},
	note = {[Accessed: 30-10-2025]},
}

@INPROCEEDINGS{10123660,
  author={Ait, Adem and Izquierdo, Javier Luis Cánovas and Cabot, Jordi},
  booktitle={2023 IEEE International Conference on Software Analysis, Evolution and Reengineering (SANER)}, 
  title={{HFCommunity: A Tool to Analyze the Hugging Face Hub Community}}, 
  year={2023},
  volume={},
  number={},
  pages={728-732},
  doi={10.1109/SANER56733.2023.00080}
}

@misc{linkedinJustCrossed,
	author = {Clem Delangue},
	title = {{W}e just crossed 1,500,000 public models on {H}ugging {F}ace --- linkedin.com},
	howpublished = {\url{https://huggingface.co/posts/clem/238420842235482/}},
	year = {},
	note = {[Accessed 29-05-2025]},
}

@book{sommerville2015,
  author    = {Ian Sommerville},
  title     = {Software Engineering},
  edition   = {10},
  year      = {2015},
  publisher = {Pearson}
}

@misc{huggingfaceHuggingFace,
  author = {{Hugging Face Inc.}},
  title = {{H}ugging {F}ace {H}ub documentation --- huggingface.co},
  howpublished = {\url{https://huggingface.co/docs/hub/index}},
  year   = {[n.d.]},
  note = {[Accessed 30-10-2025]},
}

@article{laufer2025anatomy,
  title={{Anatomy of a Machine Learning Ecosystem: 2 Million Models on Hugging Face}},
  author={Laufer, Benjamin and Oderinwale, Hamidah and Kleinberg, Jon},
  journal={arXiv preprint arXiv:2508.06811},
  year={2025}
}

@inproceedings{mitchell2019model,
  title={Model cards for model reporting},
  author={Mitchell, Margaret and Wu, Simone and Zaldivar, Andrew and Barnes, Parker and Vasserman, Lucy and Hutchinson, Ben and Spitzer, Elena and Raji, Inioluwa Deborah and Gebru, Timnit},
  booktitle={Proceedings of the conference on fairness, accountability, and transparency},
  pages={220--229},
  year={2019}
}

@misc{huggingfaceModelCardsMetadata,
	author = {{Hugging Face Inc.}},
	title = {{M}odel {C}ards - {M}odel card metadata --- huggingface.co},
	howpublished = {\url{https://huggingface.co/docs/hub/model-cards#model-card-metadata}},
	year = {},
	note = {[Accessed 30-10-2025]},
}

@misc{huggingfaceModelCardsPaper,
	author = {{Hugging Face Inc.}},
	title = {{M}odel {C}ards - {L}inking a {P}aper --- huggingface.co},
	howpublished = {\url{https://huggingface.co/docs/hub/model-cards\#linking-a-paper}},
	year = {},
	note = {[Accessed 30-10-2025]},
}

@inproceedings{suryani2024exploration,
  title={{Exploration of Hugging Face Models by Heterogeneous Information Network and Linking Across Scholarly Repositories}},
  author={Suryani, Muhammad Asif and Karmakar, Saurav and Mathiak, Brigitte},
  booktitle={International Conference on Advances in Social Networks Analysis and Mining},
  pages={371--386},
  year={2024},
  organization={Springer}
}

@misc{qualtricsSampleSize,
	author = {Qualtrics},
	title = {{S}ample {S}ize {C}alculator - {Q}ualtrics --- qualtrics.com},
	howpublished = {\url{https://www.qualtrics.com/blog/calculating-sample-size/}},
	year = {},
	note = {[Accessed 21-05-2025]},
}

@article{o2020intercoder,
  title={Intercoder reliability in qualitative research: debates and practical guidelines},
  author={O’Connor, Cliodhna and Joffe, Helene},
  journal={International journal of qualitative methods},
  volume={19},
  pages={1609406919899220},
  year={2020},
  publisher={SAGE Publications Sage CA: Los Angeles, CA}
}

@article{sim2005kappa,
  title={The kappa statistic in reliability studies: use, interpretation, and sample size requirements},
  author={Sim, Julius and Wright, Chris C},
  journal={Physical therapy},
  volume={85},
  number={3},
  pages={257--268},
  year={2005},
  publisher={Oxford University Press}
}

@article{10.1145/3695988,
author = {Hou, Xinyi and Zhao, Yanjie and Liu, Yue and Yang, Zhou and Wang, Kailong and Li, Li and Luo, Xiapu and Lo, David and Grundy, John and Wang, Haoyu},
title = {{Large Language Models for Software Engineering: A Systematic Literature Review}},
year = {2024},
publisher = {Association for Computing Machinery},
address = {New York, NY, USA},
issn = {1049-331X},
url = {https://doi.org/10.1145/3695988},
doi = {10.1145/3695988},
note = {Just Accepted},
journal = {ACM Trans. Softw. Eng. Methodol.},
month = sep,
}

@misc{githubGitHubModels,
	author = {},
	title = {{G}it{H}ub {M}odels - {G}it{H}ub {D}ocs --- docs.github.com},
	howpublished = {\url{https://docs.github.com/en/github-models}},
	year = {},
	note = {[Accessed 03-11-2025]},
}

@misc{pytorchPyTorch,
	author = {{PyTorch Foundation}},
	title = {{P}y{T}orch {H}ub --- pytorch.org},
	howpublished = {\url{https://pytorch.org/hub/}},
	year   = {[n.d.]},
	note = {[Accessed 03-11-2025]},
}

@article{han2021pre,
  title={{Pre-trained models: Past, present and future}},
  author={Han, Xu and Zhang, Zhengyan and Ding, Ning and Gu, Yuxian and Liu, Xiao and Huo, Yuqi and Qiu, Jiezhong and Yao, Yuan and Zhang, Ao and Zhang, Liang and others},
  journal={AI Open},
  volume={2},
  pages={225--250},
  year={2021},
  publisher={Elsevier}
}

@article{ait2024hfcommunity,
  title={{HFCommunity: An extraction process and relational database to analyze Hugging Face Hub data}},
  author={Ait, Adem and Izquierdo, Javier Luis C{\'a}novas and Cabot, Jordi},
  journal={Science of Computer Programming},
  volume={234},
  pages={103079},
  year={2024},
  publisher={Elsevier}
}

@inproceedings{10.1145/3643991.3644898,
author = {Casta\~{n}o, Joel and Mart\'{\i}nez-Fern\'{a}ndez, Silverio and Franch, Xavier and Bogner, Justus},
title = {Analyzing the Evolution and Maintenance of ML Models on Hugging Face},
year = {2024},
isbn = {9798400705878},
publisher = {Association for Computing Machinery},
address = {New York, NY, USA},
url = {https://doi.org/10.1145/3643991.3644898},
doi = {10.1145/3643991.3644898},
booktitle = {Proceedings of the 21st International Conference on Mining Software Repositories},
pages = {607–618},
numpages = {12},
location = {Lisbon, Portugal},
series = {MSR '24}
}

@article{giray2021software,
  title={{A software engineering perspective on engineering machine learning systems: State of the art and challenges}},
  author={Giray, G{\"o}rkem},
  journal={JSS},
  volume={180},
  pages={111031},
  year={2021},
  publisher={Elsevier}
}

@inproceedings{tan2024challenges,
  title={{Challenges of Using Pre-trained Models: the Practitioners' Perspective}},
  author={Tan, Xin and Li, Taichuan and Chen, Ruohe and Liu, Fang and Zhang, Li},
  booktitle={SANER'24},
  pages={67--78},
  year={2024},
  organization={IEEE}
}

@article{zhao2024empirical,
  title={An empirical study of challenges in machine learning asset management},
  author={Zhao, Zhimin and Chen, Yihao and Bangash, Abdul Ali and Adams, Bram and Hassan, Ahmed E},
  journal={Empirical Software Engineering},
  volume={29},
  number={4},
  pages={98},
  year={2024},
  publisher={Springer}
}

@INPROCEEDINGS{8804457,
  author={Amershi, Saleema and Begel, Andrew and Bird, Christian and DeLine, Robert and Gall, Harald and Kamar, Ece and Nagappan, Nachiappan and Nushi, Besmira and Zimmermann, Thomas},
  booktitle={2019 IEEE/ACM 41st International Conference on Software Engineering: Software Engineering in Practice (ICSE-SEIP)}, 
  title={{Software Engineering for Machine Learning: A Case Study}}, 
  year={2019},
  volume={},
  number={},
  pages={291-300},
  doi={10.1109/ICSE-SEIP.2019.00042}}

@inproceedings{10.1145/3661167.3661215, 
author = {Di Sipio, Claudio and Rubei, Riccardo and Di Rocco, Juri and Di Ruscio, Davide and Nguyen, Phuong T.},
title = {{Automated categorization of pre-trained models in software engineering: A case study with a Hugging Face dataset}},
year = {2024},
isbn = {9798400717017},
publisher = {Association for Computing Machinery},
address = {New York, NY, USA},
url = {https://doi.org/10.1145/3661167.3661215},
doi = {10.1145/3661167.3661215},
booktitle = {Proceedings of the 28th International Conference on Evaluation and Assessment in Software Engineering},
pages = {351–356},
numpages = {6},
location = {Salerno, Italy},
series = {EASE '24}
}

@inproceedings{jiang2023ptmtorrent,
  title={{Ptmtorrent: A dataset for mining open-source pre-trained model packages}},
  author={Jiang, Wenxin and Synovic, Nicholas and Jajal, Purvish and Schorlemmer, Taylor R and Tewari, Arav and Pareek, Bhavesh and Thiruvathukal, George K and Davis, James C},
  booktitle={2023 IEEE/ACM 20th International Conference on Mining Software Repositories (MSR)},
  pages={57--61},
  year={2023},
  organization={IEEE}
}

@inproceedings{jiang2024peatmoss,
  title={{Peatmoss: A dataset and initial analysis of pre-trained models in open-source software}},
  author={Jiang, Wenxin and Yasmin, Jerin and Jones, Jason and Synovic, Nicholas and Kuo, Jiashen and Bielanski, Nathaniel and Tian, Yuan and Thiruvathukal, George K and Davis, James C},
  booktitle={Proceedings of the 21st International Conference on Mining Software Repositories},
  pages={431--443},
  year={2024}
}

@inproceedings{suryani2025model,
  title={{Model Card Metadata Collection from Hugging Face to Foster Multidisciplinary AI Research: A Dataset}},
  author={Suryani, Muhammad Asif and Karmakar, Saurav and Mathiak, Brigitte and Mayr, Philipp},
  booktitle={Proceedings of the 14th International Conference on Data Science, Technology and Applications},
  pages={583--590},
  year={2025}
}

@article{koohjani2025exploring,
  title={{Exploring the Lifecycle and Maintenance Practices of Pre-Trained Models in Open-Source Software Repositories}},
  author={Koohjani, Matin and Costa, Diego Elias},
  journal={arXiv preprint arXiv:2504.06040},
  year={2025}
}

@inproceedings{jiang2023empirical,
  title={An empirical study of pre-trained model reuse in the hugging face deep learning model registry},
  author={Jiang, Wenxin and Synovic, Nicholas and Hyatt, Matt and Schorlemmer, Taylor R and Sethi, Rohan and Lu, Yung-Hsiang and Thiruvathukal, George K and Davis, James C},
  booktitle={2023 IEEE/ACM 45th International Conference on Software Engineering (ICSE)},
  pages={2463--2475},
  year={2023},
  organization={IEEE}
}

@inproceedings{davis2023reusing,
  title={{Reusing deep learning models: Challenges and directions in software engineering}},
  author={Davis, James C and Jajal, Purvish and Jiang, Wenxin and Schorlemmer, Taylor R and Synovic, Nicholas and Thiruvathukal, George K},
  booktitle={2023 IEEE John Vincent Atanasoff International Symposium on Modern Computing (JVA)},
  pages={17--30},
  year={2023},
  organization={IEEE}
}

@article{zhuang2020comprehensive,
  title={A comprehensive survey on transfer learning},
  author={Zhuang, Fuzhen and Qi, Zhiyuan and Duan, Keyu and Xi, Dongbo and Zhu, Yongchun and Zhu, Hengshu and Xiong, Hui and He, Qing},
  journal={Proceedings of the IEEE},
  volume={109},
  number={1},
  pages={43--76},
  year={2020},
  publisher={Ieee}
}

@article{zhouunifying,
  title={{A Unifying Perspective on Model Reuse: From Small to Large Pre-Trained Models}},
  author={Zhou, Da-Wei and Ye, Han-Jia}
}

@book{washizaki2024swebok,
  author    = {Hiroyasu Washizaki, eds.},
  title     = {{Guide to the Software Engineering Body of Knowledge (SWEBOK Guide), Version 4.0}},
  publisher = {IEEE Computer Society},
  year      = {2024},
  url       = {https://www.swebok.org},
}

@article{ajibode2025towards,
  title={Towards semantic versioning of open pre-trained language model releases on hugging face},
  author={Ajibode, Adekunle and Bangash, Abdul Ali and Cogo, Filipe R and Adams, Bram and Hassan, Ahmed E},
  journal={Empirical Software Engineering},
  volume={30},
  number={3},
  pages={1--63},
  year={2025},
  publisher={Springer}
}

@article{austin2021program,
  title={Program synthesis with large language models},
  author={Austin, Jacob and Odena, Augustus and Nye, Maxwell and Bosma, Maarten and Michalewski, Henryk and Dohan, David and Jiang, Ellen and Cai, Carrie and Terry, Michael and Le, Quoc and others},
  journal={arXiv preprint arXiv:2108.07732},
  year={2021}
}

@article{chen2021evaluating,
  title={Evaluating large language models trained on code},
  author={Chen, Mark},
  journal={arXiv preprint arXiv:2107.03374},
  year={2021}
}

@misc{anonymous_author_s_2025_17674909,
  author       = {Anonymous},
  title        = {{Replication Package for "SEMODS: A Validated
                   Dataset of Open-Source Software Engineering
                   Models"
                  }},
  month        = nov,
  year         = 2025,
  publisher    = {Zenodo},
  doi          = {10.5281/zenodo.17674908},
  howpublished = {\url{https://zenodo.org/records/17674909}},
}

@misc{zenodoSEMODSValidated,
	author = {Anonymous},
	title = {{S}{E}{M}{O}{D}{S}: {A} {V}alidated {D}ataset of {O}pen-{S}ource {S}oftware {E}ngineering {M}odels ({N}ovember 2025 {S}napshot) --- zenodo.org},
    doi = {10.5281/zenodo.17675255},
	howpublished = {\url{https://zenodo.org/records/17675256}},
	month        = nov,
    year         = 2025,
}

\end{document}